\documentclass[preprint,review,12pt]{elsarticle}

\usepackage{graphicx,color}

\usepackage{amssymb}

\usepackage{amsmath}

\journal{Nuclear Instruments and Methods in Physics Research A}

\begin{document}
\begin{frontmatter}



\title{Development of a Pixel Detector for Ultra-Cold Neutrons}


\author[Tokyo]{S.~Kawasaki\corref{cor1}}
\ead{kawasaki@icepp.e.u-tokyo.ac.jp}

\author[Tokyo]{G.~Ichikawa}
\author[KUR]{M.~Hino}
\author[Icepp]{Y.~Kamiya}
\author[KUR]{M.~Kitaguchi}
\author[Tokyo,Icepp]{S.~Komamiya}
\author[Tohoku]{T.~Sanuki}
\author[Tokyo]{S.~Sonoda}

\address[Tokyo]{Graduate School of Science, University of Tokyo, Bunkyo-ku, Tokyo 113-0033, Japan}
\address[Icepp]{International Center for Elementary Particle Physics, University of Tokyo, Bunkyo-ku, Tokyo 113-0033, Japan}
\address[Tohoku]{Graduate School of Science, Tohoku University, Sendai-shi, Miyagi 980-8578, Japan}
\address[KUR]{Research Reactor Institute, Kyoto University, Sennan-gun, Osaka 590-0494, Japan}

\cortext[cor1]{Corresponding author: Tel: +81-3-5841-4201; Fax: +81-3-5841-4158}

\begin{abstract}
A pixel detector with high spatial resolution and temporal information for ultra-cold neutrons is developed
based on a commercial CCD on which a neutron converter is attached.
$\mathrm{^{10}B}$ and $\mathrm{^6Li}$ are tested for the neutron converter
and $\mathrm{^{10}B}$ is found to be more suitable based on efficiency and spatial resolution.
The pixel detector has an efficiency of 44.1$\pm$1.1\% and a spatial resolution of 2.9$\pm$0.1 $\mathrm{\mu m}$ (1 sigma).

\end{abstract}


\end{frontmatter}


\section{Introduction}
When ultra-cold neutrons (UCNs) are trapped in the earth's gravitational field, their energy is quantized.
In consequence, their probability density distribution exhibits their vertical modulation.
The scale of this density modulation is calculated to be $(\hbar^2/2m_n^2g)^{1/3} \sim 6 \ \mathrm{\mu m}$. 
Observations of such quantum states have been presented in Refs. \cite{bib_Nesvizhevsky2002na,bib_Nesvizhevsky2003pr,bib_Nesvizhevsky2005ep}.
We have proposed a more precise measurement using a pixel detector with an image magnification system \cite{bib_Sanuki}.
Another pixel detectors for UCN have been developed recently, such as that reported in Ref. \cite{bib_timepix}.
In this article we present the development of a pixel detector based on a commercial charge coupled device (CCD) covered with a neutron converter.
By comparing $\mathrm{^{10}B}$ and $\mathrm{^6Li}$ as converter materials,
we find that a $\mathrm{^6Li}$ converter produces energetic tritons which penetrate deep into the CCD in various directions, 
degrading the spatial resolution.
Hence we conclude that $\mathrm{^{10}B}$ is an appropriate material for a neutron converter.

\section{Detector design}
The developed detector consists essentially of a CCD covered by a neutron converter. Charged particles produced via nuclear reaction in the converter are detected with the CCD.
The choices of converter material and CCD are key for this detector.

\subsection{Neutron converter}
$\mathrm{^{10}B}$ and $\mathrm{^6Li}$ are chosen as test materials for the neutron converter, because of their large cross-sections with neutrons.
The neutron absorption cross-sections for $\mathrm{^{10}B}$ and $\mathrm{^6Li}$ are $4.01 \times 10^3$ and $0.95 \times 10^3$ barn, respectively, for thermal neutrons ($v$ = 2,224 m/s).

Neutrons react with $\mathrm{^{10}B}$ and $\mathrm{^6Li}$ in the following processes:
\begin{align}
\mathrm{n + ^{10}B }& \rightarrow \mathrm{\alpha\ (1.47\ MeV) + ^7Li\ (0.84\ MeV) + \gamma\ (0.48\ MeV)} &\mathrm{93.9\%}  \label{eq_converterB}\\
				    & \rightarrow \mathrm{\alpha\ (1.78\ MeV) + ^7Li\ (1.01\ MeV)} & \mathrm\ {6.1\%} \notag \\
\mathrm{n + ^6Li}   & \rightarrow \mathrm{\alpha\ (2.05\ MeV) + \mathrm{^3H}\ (2.73\ MeV)}& \mathrm{100\%}
\label{eq_converterLi}
\end{align}

$\mathrm{^{10}B}$ has advantages over $\mathrm{^6Li}$ in that it has a larger cross section and a shorter range for converted charged particles in the CCD;
$\mathrm{^{10}B}$ emits only short range particles.
The ranges of $\alpha$(1.47 MeV), $\alpha$(1.78 MeV) and $\mathrm{^7Li}$(0.84 MeV, 1.01 MeV) in Si are 5, 6 and 2 $\mu$m, respectively. 
In contrast, $\mathrm{^6Li}$ emits $\alpha$ particles and tritons with ranges of 7 and 40 $\mu$m.
Long range tritons degrade the spatial resolution, as discussed in Ref. \cite{bib_Sanuki}.

\subsection{CCD sensor}
A CCD is an ideal device for UCN detection because of its high spatial resolution and its ability to capture data in real time. 
Since a CCD cannot directly detect neutral particles, a neutron converter to create charged particles must be attached to the front of the device.
A back-thinned type CCD is chosen to avoid a large insensitive volume in front of the active volume of the CCD, which would prevent converted charged particles entering the active volume or degrade the spatial resolution.
We use a commercial CCD detector, HAMAMATSU S7170-0909, in the low dark current read out mode.
The specifications are shown in Table \ref{CCD:spec}.
Generally, CCD might have white-spot pixels induced by radiation damage.  
Only 2 pixels have become noisy after 60 million neutron irradiation per 512$\times$512 pixels during the neutron beam tests mentioned below. 
\begin{table}[htdp]
\caption{Specifications of the HAMAMATSU S7170-0909 ($\mathrm{T_a = 25}$\kern-.2em\r{}\kern-.3em C)}
\begin{center}
\begin{tabular}{ll}
\hline
Parameter & \\
\hline
Active Area	&
12.288 $\times$ 12.288 mm
\\
Number of Pixels	&
512 $\times$ 512
\\
Pixel Size	&
24 $\times$ 24 $\mu$m
\\
Frame Rate	&
0.9 frames/s
\\
Full Well Capacity (Vertical)	&
300 k$e^-$
\\
Full Well Capacity (Horizontal)	&
600 k$e^-$
\\
Dark Current Max. $\mathrm{0^\circ C}$	&
600 $e^-$/pixel/s
\\
Readout Noise	&
8 $e^-$rms																											
\\
\hline
\end{tabular}
\end{center}
\label{CCD:spec}
\end{table}

\subsection{Fabrication of a converter layer}
The neutron converter is attached directly to the surface of a CCD to minimize the distance between the positions of an incident neutron and a detected charged particle \cite{bib_Sanuki}.
The converter layer is mounted on the CCD surface by vacuum evaporation at Kyoto University Research Reactor Institute.
The pressure of the evaporation chamber is around $\mathrm{10^{-3} \ Pa}$ and the evaporation rate is about 1 \AA $\mathrm{/s}$ for each material.
The facility used is described in Ref. \cite{bib_KURRI}.

Converter layers of 46 $\mathrm{\mu g \,cm^{-2}}$ ($\mathrm{220\ nm}$) and 11 $\mathrm{\mu g \, cm^{-2}}$ ($\mathrm{230\ nm}$) for $\mathrm{^{10}B}$ and $\mathrm{^6Li}$, respectively, are attached to the CCD surface.
To form a firm layer, a 9 $\mathrm{\mu g \, cm^{-2}}$(20\ nm) Ti layer is directly deposited on the CCD surface and the converter layer is mounted on the Ti layer.
To repel moisture from the air, the outer surface of the converter layer is covered with a 9 $\mathrm{\mu g \, cm^{-2}}$(20\ nm) Ti layer.
(Note that we incorrectly reported the amount of Ti and Li layer in our previous report \cite{bib_Sanuki}.)
Ti has good characteristics for this use, because of its chemical stability, adhesion and negative potential for neutrons.
The stability of these Ti--$\mathrm{^{10}B}$--Ti and Ti--$\mathrm{^6Li}$--Ti structures have allowed our developed detector to work reliably for more than three years.

\section{Performance of a CCD sensor}
\label{pef.:CCD}
We investigate the CCD response to charged particles using  $\alpha$ particles from $\mathrm{^{241}Am}$.
The CCD and the $\mathrm{\alpha}$ source are put in a chamber filled with dry $\mathrm{N_2}$ gas, with the distance between them fixed at $\mathrm{123\ mm}$.
The $\mathrm{N_2}$ gas pressure is varied to change the energy of the $\mathrm{\alpha}$ particles at the surface of the CCD.
The energy to create an electron--hole pair in Si is $\mathrm{3.65\ eV}$, and hence an energy deposit of $\mathrm{1\ MeV}$, for instance, creates about 300,000 electron--hole pairs.

\begin{figure}[htdp]
\begin{center}
	\resizebox{80mm}{!}{\includegraphics{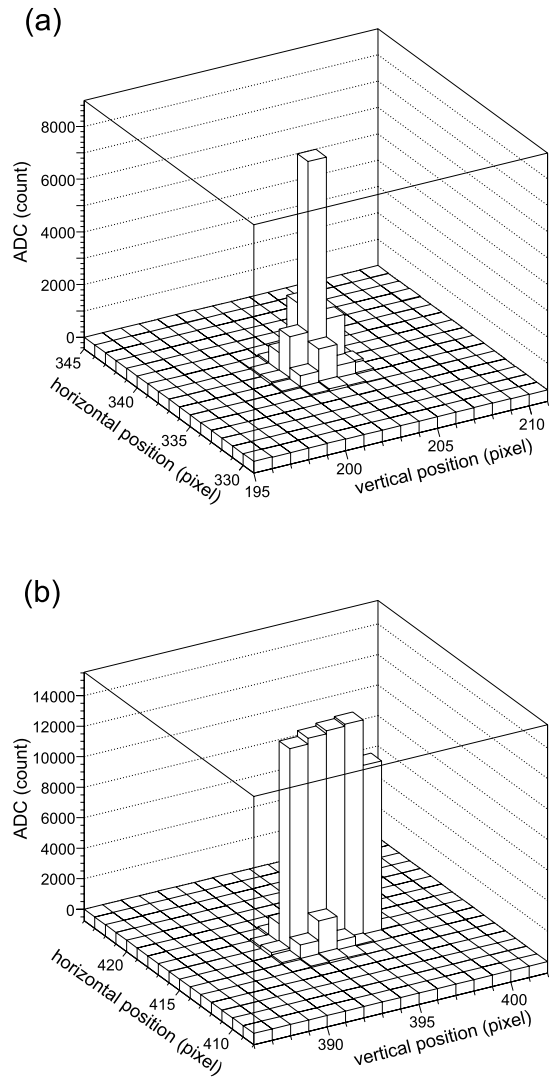}}
	\caption{Typical signals made by $\alpha$ particles of energy (a) 1 MeV and (b) 4 MeV. An $\alpha$ particle of energy lower than 2 MeV creates an isotropic cluster, while an $\alpha$ particle of energy higher than 2 MeV creates an anisotropic cluster spread in the vertical direction.}
	\label{signals}
\end{center}
\end{figure}

Electrons are collected at the anode and stored in a charge capacitance.
During storage, electrons spread to the adjacent pixels and make a cluster.
Fig. \ref{signals} shows typical cluster shapes made by $\mathrm{\alpha}$ particles of energy 1 MeV and 4 MeV.
For $\alpha$ particles with energy lower than 2 MeV, the created electrons diffuse isotropically in the vertical and horizontal directions.
For more energetic particles, higher than 2 MeV, the cluster shapes become anisotropic.
This is because so many electron--hole pairs are created that they overflow the full well capacity. 
As shown in Table \ref{CCD:spec}, the vertical full well capacity is lower than the horizontal capacity.
Overflowed electrons spread only in the vertical direction and make an anisotropic cluster.

\begin{figure}[tdp]
\begin{center}
	\resizebox{100mm}{!}{\includegraphics{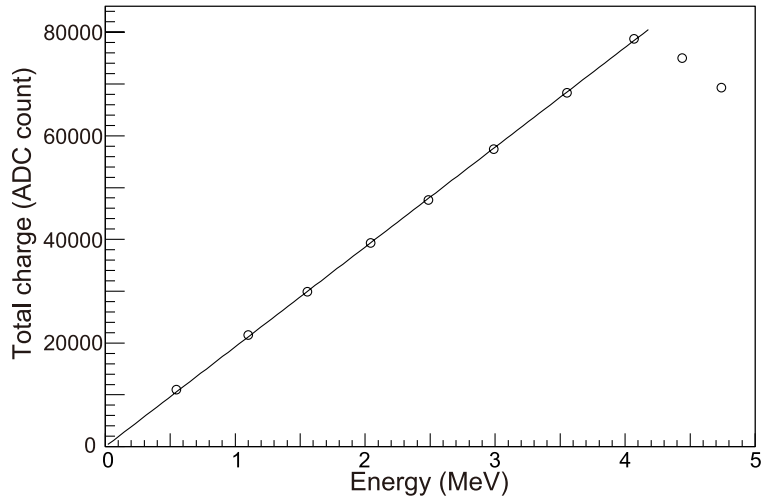}}
	\caption{Energy calibration curve. The total charge in the signal region is proportional to the energy of the $\alpha$ particle below 4 MeV.}
	\label{calibration}
\end{center}
\end{figure}

Considering the anisotropy of the full well capacity, the signal region is determined to be  $\mathrm{7 \ pixels}$ (horizontal) $\times $ $\mathrm{11 \ pixels}$ (vertical) around the peak pixel, collecting all the created electrons.
More than 99\% of the diffused electrons are stored in this region. 
Fig. \ref{calibration} shows the energy calibration curve of the total charge in the signal region to the energy of $\alpha$ particles.
Linearity is confirmed below 4 MeV.
$\alpha$ particles of energy more than 4 MeV are so energetic that they can penetrate a CCD detection volume thickness of 20 $\mathrm{\mu m}$.

When the neutron rate is 100 events/frame, more than one event is observed inside one signal region.

\section{Neutron beam test}
The performance of neutron detectors with a $\mathrm{^{10}B}$ converter and a $\mathrm{^6Li}$ converter are tested with neutrons from reactors.

\subsection{Neutron detection efficiency}
\begin{figure}[htdp]
\begin{center}
	\resizebox{100mm}{!}{\includegraphics{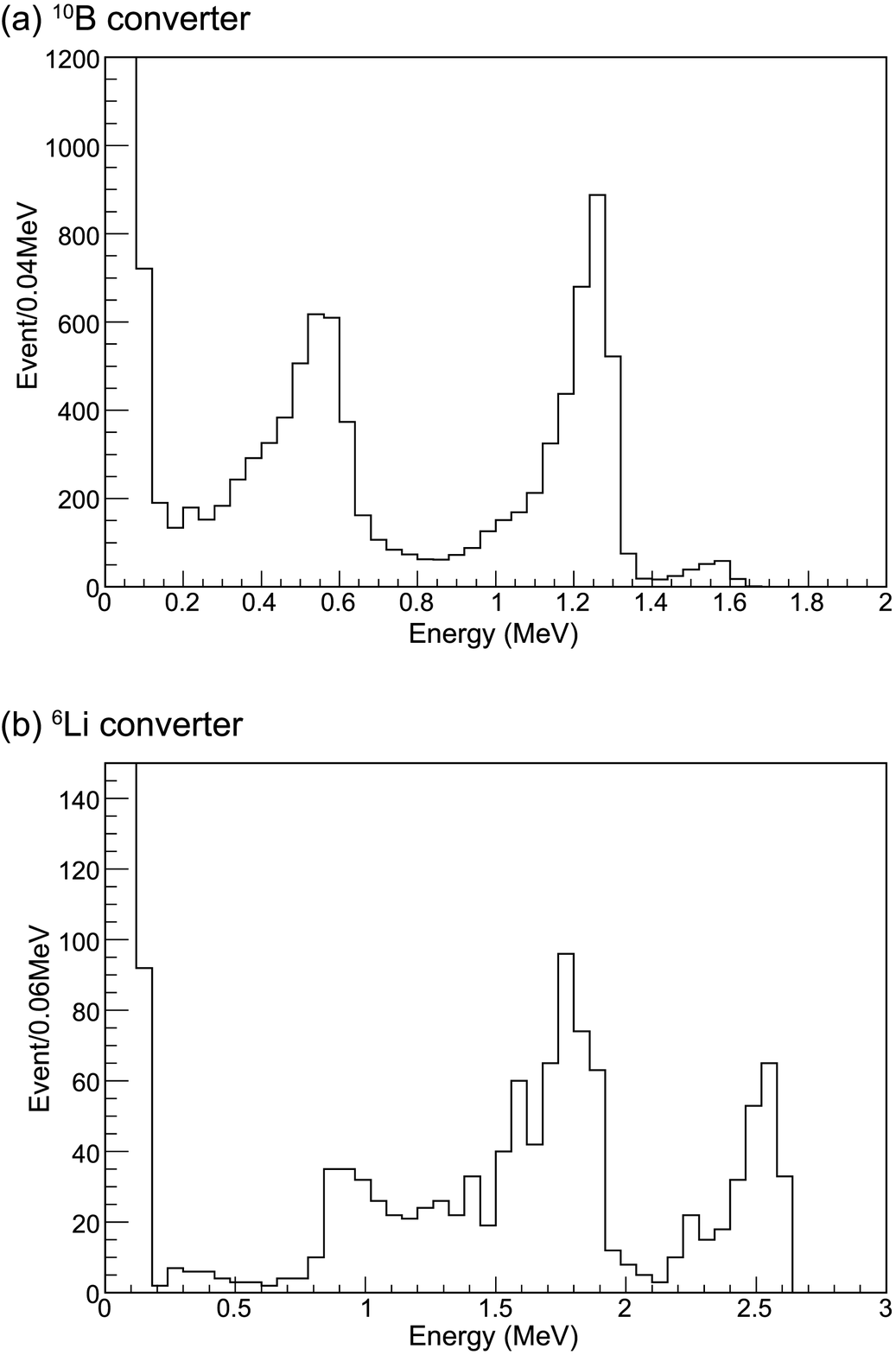}}
	\caption{Energy spectra observed with CCDs with (a) $\mathrm{^{10}B}$ converter and (b) $\mathrm{^6Li}$ converter}
	\label{energy}
\end{center}
\end{figure}

The neutron detection efficiency is measured using UCNs supplied at the PF2 beam line of Institut Laue--Langevin in France.
Fig. \ref{energy} shows the energy spectra measured by CCDs with $\mathrm{^{10}B}$ and $\mathrm{^6 Li}$ converters.
Peaks corresponding to the converted charged particles (see Eq. \ref{eq_converterB}, \ref{eq_converterLi}) are seen in the spectra.
The peak energy is slightly lower than the converted charged particle energy because of the energy loss in the converter itself.
A peak around $\mathrm{0.9\ MeV}$ in the energy spectrum for the $^6 \mathrm{Li}$ converter is caused by tritons which are so energetic that they penetrate through the sensitive volume of the CCD \cite{bib_Sanuki}.
Tritons emitted in a large incident angle at the surface of the CCD deposit all their energy and stop in the CCD.

To select neutron events, the minimum energy cut is set to 0.3 MeV for the $^{10} \mathrm{B}$ converter and to 0.8 MeV for the $^6 \mathrm{Li}$ converter.
The incident neutron flux is measured with a $^3\mathrm{He}$ gas counter with an efficiency greater than 83\%.
The detection efficiency is measured to be 44.1\,$\pm$\,0.7(stat.)\,$\pm$\,0.8(syst.)\% for the $^{10} \mathrm{B}$ converter and 21.0\,$\pm$\,0.8(stat.)\,$\pm$\,0.5(syst.)\% for the $^6 \mathrm{Li}$ converter, normalized by the number of neutrons measured by the $\mathrm{^3He}$ gas counter.
The main source of the systematic error is uncertainty in the initial flux of neutrons.
The neutron conversion efficiency for the $\mathrm{^{10}B}$ converter is higher than that for the $\mathrm{^6Li}$ as expected from their cross-sections.

\subsection{Spatial resolution}
The electrons spread over nearby pixels and make a cluster, as shown in Fig. \ref{signals}.
The barycenter of charge is a good estimation of the neutron incident position.

We measure the spatial resolution of the CCD using reference patterns made with a neutron absorber.
The reference pattern is placed just in front of the CCD and is irradiated by neutrons.
The distance between the reference pattern and the CCD surface is as close as $\mathrm{150\ \mu m}$.

The reference pattern and the CCD are exposed to cold neutrons (CNs) at the MINE2 beam line of the research reactor JRR-3. 
The monochromatic neutron beam has a wavelength of 0.88 nm and a bandwidth of 2.7\% in full width at half maximum.
CNs are suitable for evaluating the spatial resolution because their beam divergence is as small as $1\times10^{-3}$ radian.

\subsubsection{Comparison between $\mathrm{^{10}B}$ and $\mathrm{^6Li}$ converters} 
We have reported that a CCD sensor covered with a $\mathrm{^6Li}$ converter can determine a neutron incident position with a precision of $\mathrm{\mathrm{5.3\ \mu m}}$ \cite{bib_Sanuki}.
The spatial resolution of a CCD with a $\mathrm{^{10}B}$ converter is compared to that of a $\mathrm{^6Li}$ converter
using the same reference pattern as used in our previous report.

\begin{figure}[htdp]
\begin{center}
	\resizebox{100mm}{!}{\includegraphics{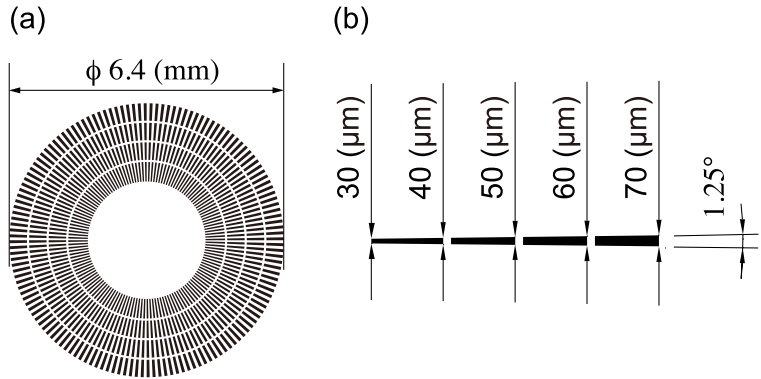}}
	\caption{(a) Reference pattern created by a laser process. (b) Enlargement of one spoke. The black areas are hole positions.}
	\label{laser_pattern}
\end{center}
\end{figure}

The reference pattern is made of a gadolinium foil with a thickness of $\mathrm{25\ \mu m}$.
This thickness is sufficient to fully absorb CNs.
The 144 fine spoke slits shown in Fig. \ref{laser_pattern}(a) are formed by an excimer laser by Laserx Co. Ltd.
Each spoke slit has four trapezoidal holes along the radius, as shown in Fig. \ref{laser_pattern}(b) \cite{bib_Sanuki}.

The spatial resolution depends on the range and deposit energy of the charged particle created by nuclear reaction in the converter.
To investigate differences in the spatial resolution of charged particles, the energy range is selected to identify different charged particles (see Table \ref{different_resolution}).

\begin{figure}[htdp]
\begin{center}
	\resizebox{80mm}{!}{\includegraphics{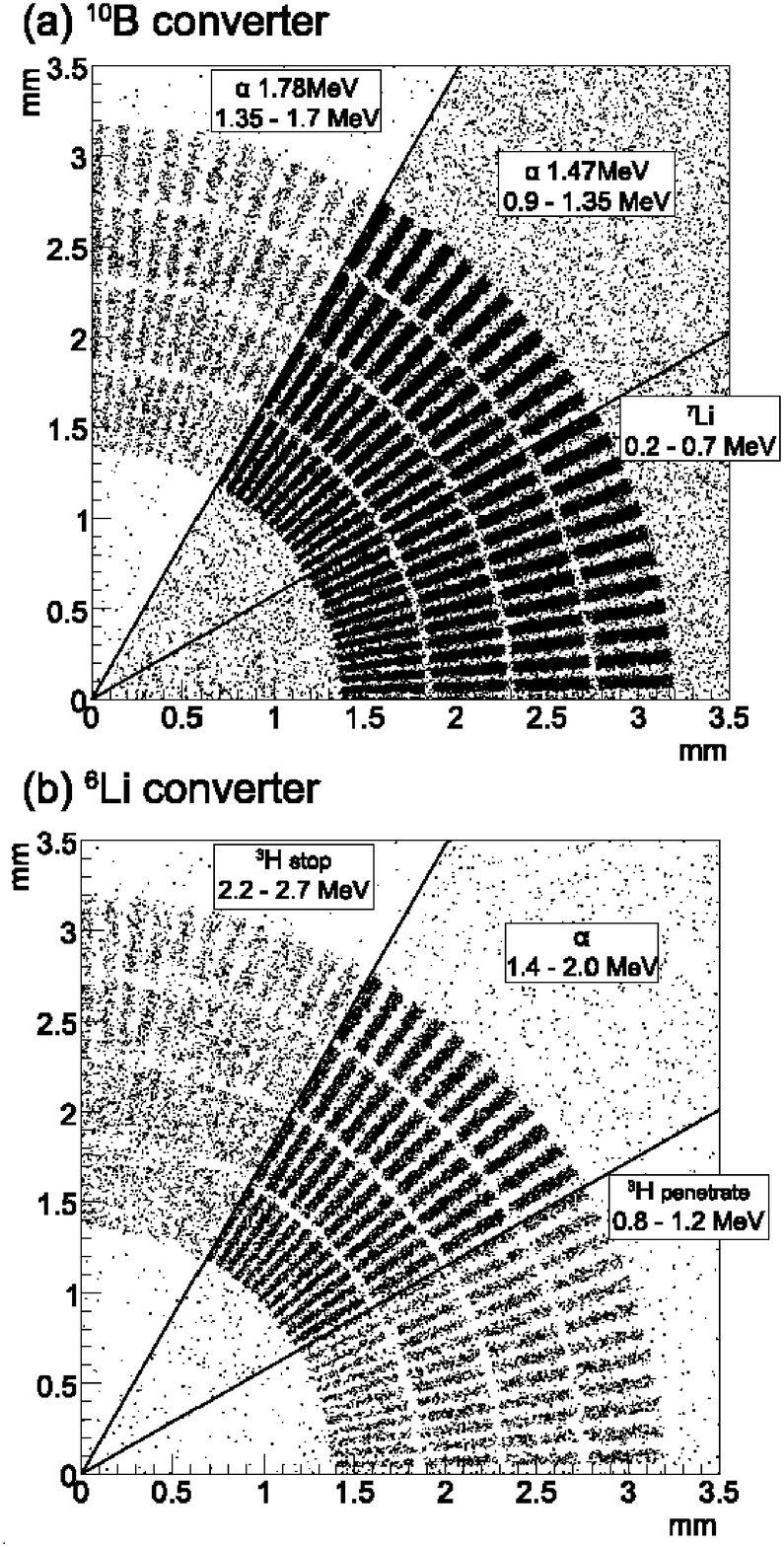}}
	\caption{Hit position distribution measured with a pattern created by a laser process}
	\label{scat}
\end{center}
\end{figure}

Fig. \ref{scat} shows the hit position distributions calculated from the barycenter of each cluster.
All the events are plotted in an area one-twelfth of the total area, assuming rotational symmetry of the event distribution.
For the stopped tritons from the $^6 \mathrm{Li}$ converter, the innermost slit cannot be distinguished.
This is because the stopped tritons span a long range inside the CCD, as mentioned above.
On the other hand, a clear pattern can be seen for other charged particles.
These figure indicate that the spatial resolution is much better than the smallest pattern distance of $\mathrm{30 \ \mu m}$.

\begin{figure}[htdp]
\begin{center}
\resizebox{100mm}{!}{\includegraphics{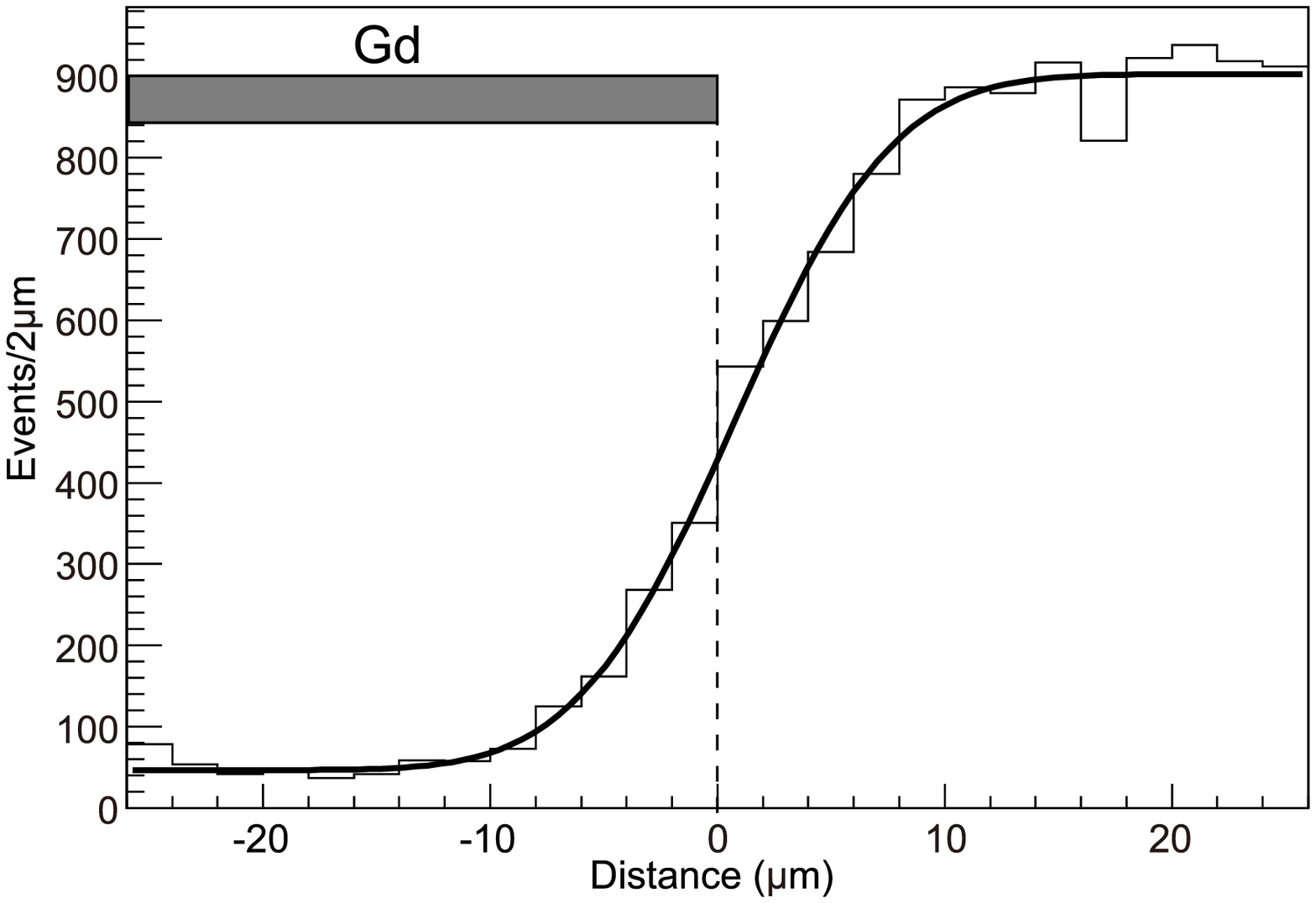}} 
	\caption{Distance between the hit position and the edge of slit. Only $\alpha$(1.47 MeV) events from the $\mathrm{^{10}B}$ converter are displayed.}
	\label{fit}
\end{center}
\end{figure}

In order to estimate the spatial resolution more precisely, the hit position distribution at the edge of the slit is investigated.
The spatial resolution is evaluated by fitting an error function with a constant background expressed as
\begin{equation}
	f(x) = p_1 \cdot \mathrm{erf} \left(\frac{x}{\sqrt{2} \sigma} +p_2 \right) + p_3,
\end{equation}
where $x$ is the distance from the edge, $\sigma$ corresponds to the spatial resolution and $p_1$, $p_2$ and $p_3$ are free fitting parameters.
The fitting result for $\alpha$(1.47 MeV) from the $\mathrm{^{10}B}$ converter is shown in Fig. \ref{fit} as an example.
The obtained resolutions ($\sigma$ of the error function) are shown in Table \ref{different_resolution}.
The spatial resolution of the stopped tritons is too poor to be evaluated by the fitting.

We find that the $\mathrm{^{10}B}$ converter has higher efficiency and better spatial resolution than the $\mathrm{^6Li}$ converter.
We conclude that $\mathrm{^{10}B}$ is a more suitable converter material for a UCN detector.

\begin{table}
\caption{resolution of each charged particle}
\label{different_resolution}
\begin{center}
\begin{tabular}{cccc}
\hline
 converter &energy range (MeV)&particle& $\sigma$ ($\mathrm{\mu m}$) \\ \hline
				& 0.2 -- 0.7		& $\mathrm{^7 Li}$		& 5.7 $\pm$ 0.2\\ 
$\mathrm{^{10}B}$	& 0.9 -- 1.35	& $\mathrm{ \alpha}$	& 5.5 $\pm$ 0.2 \\
				& 1.35 -- 1.7	& $\mathrm{ \alpha}$	& 7.6 $\pm$ 1.0 \\
\hline
				& 0.8 -- 1.2 	&$\mathrm{^3He (penetrate)}$& 6.7 $\pm$ 0.6  \\
$\mathrm{^6Li}$	& 1.4 -- 2.0 	& $\mathrm{\alpha}$		& 5.3 $\pm$ 0.3 \\
				& 2.2 -- 2.7		& $\mathrm{^3He (stop)}	$	&  --- \\ \hline
\end{tabular}
\end{center}
\end{table} 

\subsubsection{Precise evaluation of spatial resolution}
The reference pattern used in the above test does not have perfectly straight edges,
probably because the gadolinium foil is too thick to form sharp edges.
The variations from a straight slit edge are observed to be 1--2 $\mathrm{\mu m}$ by optical microscopy.
The spatial resolution evaluated using the reference pattern (Table \ref{different_resolution}) includes this effect \cite{bib_Sanuki}.

\begin{figure}[thbp]
\begin{center}
	\resizebox{100mm}{!}{\includegraphics{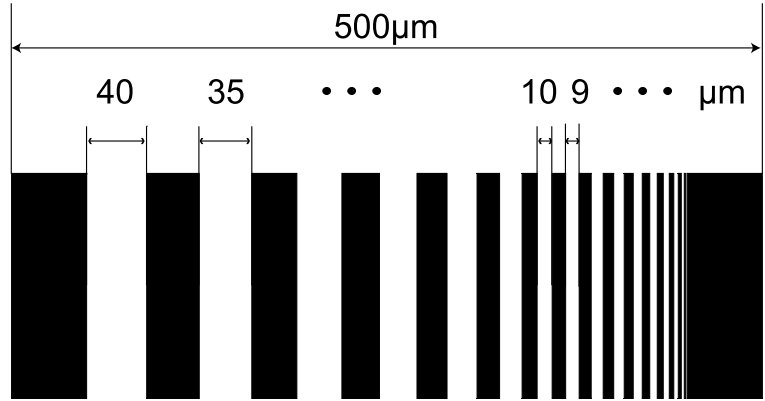}}
	\caption{Reference pattern created by damascene process. The black areas are Gd-less areas.}
	\label{Gd_pattern}
\end{center}
\end{figure}

\begin{figure}[bhtp]
\begin{center}
	\resizebox{100mm}{!}{\includegraphics{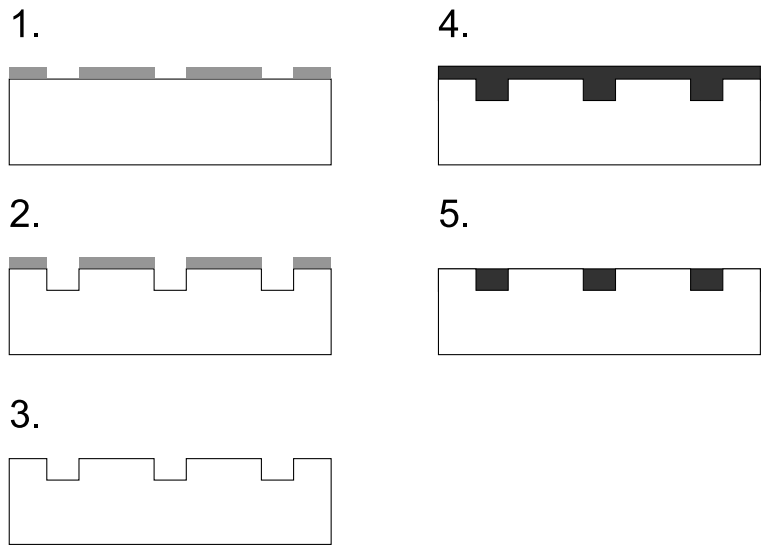}}
	\caption{Damascene process:
	1 depositing a hard mask layer on a Si substance by photo lithography,
	2 making trenches on the Si with plasma reactive ion etching,
	3 removing the mask layer,
	4 sputtering gadolinium onto the trenches,
	5 removing excess gadolinium by chemical-mechanical polishing
	}
	\label{damascene}
\end{center}
\end{figure}

To evaluate the intrinsic spatial resolution in our detector, we fabricated a more precise reference pattern
at the Center for Integrated Nano Technology Support of Tohoku University.
Fig. \ref{Gd_pattern} shows the layout of the reference Gd pattern.
The pattern consists of stripes that have different widths of 1--40 $\mathrm{\mu m}$.
The pattern is fabricated by a damascene process, usually used for semiconductor devices \cite{bib_damascene}.
The damascene process is shown in Fig. \ref{damascene}

\begin{figure}[htdp]
\begin{center}
	\resizebox{100mm}{!}{\includegraphics{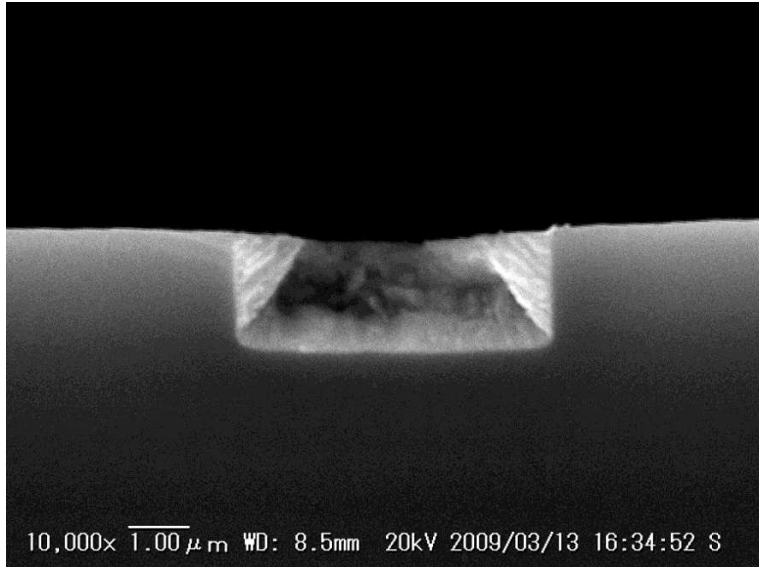}}
	\caption{A 5 $\mu$m-wide trench made by damascene process}
	\label{5um_trench}
\end{center}
\end{figure}

This process produces an edge to the pattern that is sufficiently sharp.
The thickness of the gadolinium is 2--5 $\mathrm{\mu m}$, which corresponds to an absorption probability of 
 75--97\% for CNs.
Fig. \ref{5um_trench} shows the 5 $\mu$m-wide trench after polishing.

\begin{figure}[htdp]
\begin{center}
	\resizebox{80mm}{!}{\includegraphics{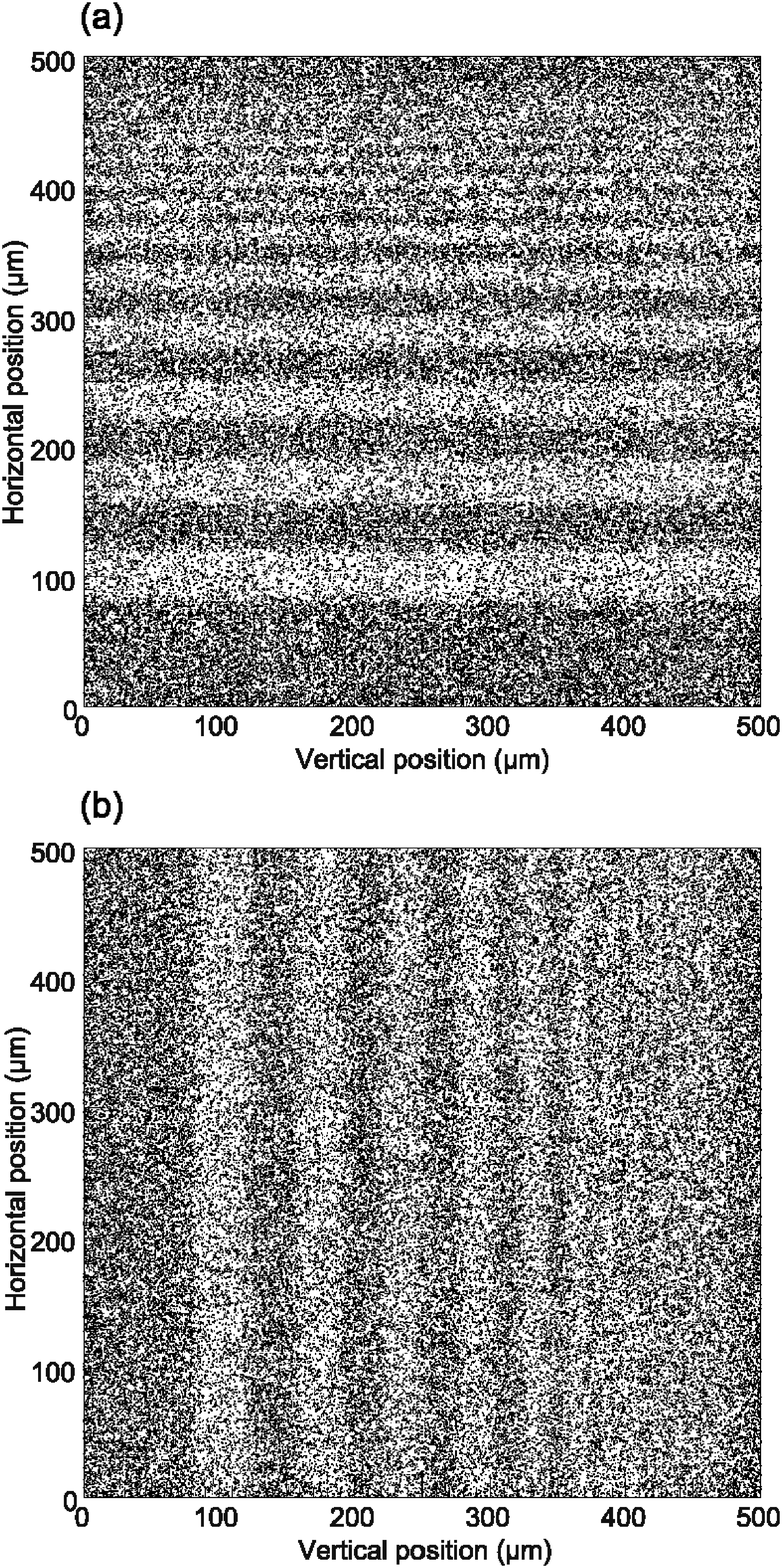}}
	\caption{Hit position distribution measured with a pattern created by a damascene process. The striped pattern is placed along the vertical direction (a) and the horizontal direction (b) }
	\label{scatterHV}
\end{center}
\end{figure}

\begin{figure}[htdp]
\begin{center}
	\resizebox{100mm}{!}{\includegraphics{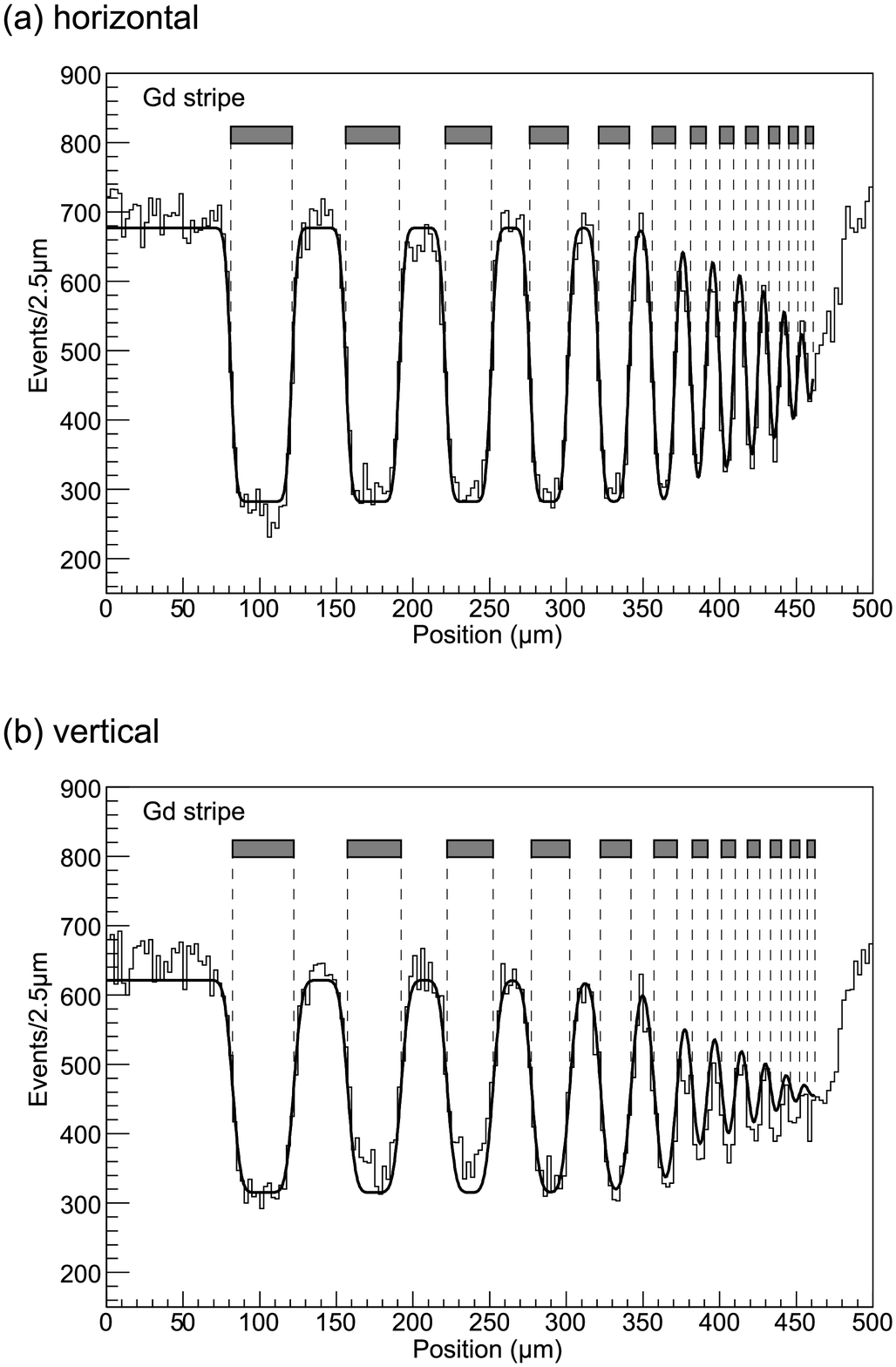}}
	\caption{Projected hit position: (a) horizontal projection of Fig.\ref{scatterHV}(a) and  (b) vertical projection of Fig.\ref{scatterHV}(b). The spatial resolutions are evaluated to be 2.9\,$\pm$\,0.1 $\mathrm{\mu m}$ (horizontal) and 4.3\,$\pm$\,0.2 $\mathrm{\mu m}$ (vertical) in terms of sigma of error function}
	\label{resolution}
\end{center}
\end{figure}

Using this striped pattern, the differences in the spatial resolution with respect to the read out directions are measured.
Fig. \ref{scatterHV} shows the hit position distribution measured with the pattern, and Fig. \ref{resolution} shows the their projections onto the horizontal and vertical direction.

The spatial resolution is evaluated by fitting.
The fitting function is expressed by the sum of an error function and a complementary error function:
\begin{equation}
 f(x) = p_1\cdot \sum_{n=1}^{12} \left[ \mathrm{erfc} \left(\frac{x-d_n}{\sqrt{2} \sigma}+p_2 \right) + \mathrm{erf} \left(\frac{x-(d_n+w_n)}{\sqrt{2} \sigma}+p_2 \right) \right] + p_3,
\end{equation}
where $x$ is the hit position, $\sigma$ is the spatial resolution, $d_n$ is the distance from the initial position of the pattern to the  $n_{th}$ stripe, $w_n$ is the width of the $n_{th}$ stripe and $p_1$, $p_2$ and $p_3$ are free fitting parameters.
The 12 widest stripes, whose widths are 5--40 $\mathrm{\mu m}$, are used for this fitting.
This is because gadolinium cannot enter trenches narrower than $\mathrm{5\ \mu m}$ in the spattering process.

In the fitting result, the horizontal and vertical resolutions are evaluated to be 2.9\,$\pm$\,0.1 $\mathrm{\mu m}$ and 4.3\,$\pm$\,0.2 $\mathrm{\mu m}$, respectively.
These values correspond to 6.9\,$\pm$\,0.2 $\mathrm{\mu m}$ and 9.9\,$\pm$\,0.3 $\mathrm{\mu m}$ of the full width at half maximum of the line spread functions.
This difference is caused by the difference in the full well capacity.
The measured spatial resolution is much finer than the pixel size of 24 $\mathrm{\mu m}$.

\section{Summary}
We developed a pixel detector for UCN which has high spatial resolution with temporal information.
The detector is based on a commercial CCD on which a suitable neutron converter is directly evaporated.
$\mathrm{^{10}B}$ and $^6\mathrm{Li}$ are tested as neutron converter materials.
We conclude that $\mathrm{^{10}B}$ has better properties as a neutron converter material.
Using $\mathrm{^{10}B}$, the UCN detection efficiency is 44.1\,$\pm$\,1.1\%.
Our UCN detector can measure the UCN incident position with a precision of 2.9\,$\pm$\,0.1 $\mathrm{\mu m}$ (horizontal) and 4.3\,$\pm$\,0.2 $\mathrm{\mu m}$ (vertical) in terms of sigma of the error functions.

\section{Acknowledgment}
We are indebted to Prof. H. M. Shimizu of the High Energy Accelerator Research Organization (KEK) for constructive comments and encouragement.
We would also like to thank Prof. P. Geltenbort of Institut Laue--Langevin (ILL)  and Prof. S. Tanaka of Tohoku University for their technical support and encouragement.
A part of this study was supported by the Center for Integrated Nanotechnology Support at Tohoku University and also by Nanotechnology Network Project of the Ministry of Education, Culture, Sports, Science, and Technology (MEXT) of the Japanese Government. 
This study was also supported by a Grant-in-Aid for Scientific Research (KAKENHI: 18409056, 20340050) from the Japan Society for the Promotion of Science and Research Fellowships of the Japan Society for the Promotion of Science for Young Scientists (19.4404).



\end{document}